\documentclass[twocolumn,secnumarabic,amssymb, aps,prb,superscriptaddress,floatfix]{revtex4-1}
\usepackage{graphicx,color,dcolumn,bm}
\usepackage{perpage}

\MakePerPage{footnote}

\usepackage[dvipdfm, pdfstartview=FitH, CJKbookmarks=true, bookmarksnumbered=true, bookmarksopen=true, colorlinks, linkcolor=blue, anchorcolor=blue, citecolor=blue]{hyperref}

\usepackage{multirow}

\begin{document}

\
\title{The interplay of electronic reconstructions, lattice distortions, and surface oxygen vacancies in  insulator-metal transition of LaAlO$_{3}$/SrTiO$_{3}$}

\author {Jun Zhou}
\affiliation{NUSNNI-Nanocore, National University of Singapore, Singapore 117542}
\affiliation{Department of Physics, National University of Singapore, Singapore 117542}
\affiliation{Singapore Synchrotron Light Source, National University of Singapore, Singapore 117603}

\author{Teguh Citra Asmara}
\affiliation{NUSNNI-Nanocore, National University of Singapore, Singapore 117542}
\affiliation{Department of Physics, National University of Singapore, Singapore 117542}
\affiliation{Singapore Synchrotron Light Source, National University of Singapore, Singapore 117603}

\author {Ming Yang}
\affiliation{NUSNNI-Nanocore, National University of Singapore, Singapore 117542}
\affiliation{Department of Physics, National University of Singapore, Singapore 117542}
\affiliation{Singapore Synchrotron Light Source, National University of Singapore, Singapore 117603}

\author{George A. Sawatzky}
\affiliation{Department of Physics and Astronomy, University of British Columbia, Vancouver, Canada V6T1Z1}

\author{Yuan Ping Feng}
\affiliation{Department of Physics, National University of Singapore, Singapore 117542}

\author {Andrivo Rusydi}
\email {phyandri@nus.edu.sg}
\affiliation{NUSNNI-Nanocore, National University of Singapore, Singapore 117542}
\affiliation{Department of Physics, National University of Singapore, Singapore 117542}
\affiliation{Singapore Synchrotron Light Source, National University of Singapore, Singapore 117603}

\date{\today}
\begin{abstract}

The mechanism responsible for the extraordinary interface conductivity of LaAlO$_{3}$ on SrTiO$_{3}$ and its insulator-metal transition remains controversial. Here, using density functional theory calculations, we establish a comprehensive and coherent picture that the interplay of electronic reconstructions, lattice distortions, and surface oxygen vacancies fully compensates the polarization potential divergence in LaAlO$_{3}$/SrTiO$_{3}$, explaining naturally the experimental observations under different conditions. While lattice distortions and a charge redistribution between LaO and AlO$_2$ sub-layers play a dominant role in insulating state, a spontaneous appearance of 1/4 oxygen vacancies per AlO$_{2}$ sub-layer at the LaAlO$_{3}$ surface accompanied by 0.5$e^{-}$ charge-transfer into the interface is responsible for interface conductivity and the discontinuous transition in LaAlO$_{3}$/SrTiO$_{3}$. Our model also explain properties of LaAlO$_{3}$/SrTiO$_{3}$ prepared with different growth conditions.

\end{abstract}
\maketitle
\section {INTRODUCTION}

The interface of dissimilar oxide materials hosts rich varieties of exotic phenomena not found in its constituent materials and has attracted a tremendous amount of research interests, both for fundamental physics and practical applications\cite{a1,a10,a11,a12,a20}. An ideal example is the interface of two different insulators, polar LaAlO$_{3}$ on nonpolar SrTiO$_{3}$ (LaAlO$_{3}$/SrTiO$_{3}$). In their bulk forms, both LaAlO$_{3}$ and SrTiO$_{3}$ are wide band gap non-magnetic insulators\cite{a13}. Remarkably, when a thin film of LaAlO$_{3}$ was deposited on SrTiO$_{3}$, the interface was found to exhibit unusual phenomena such as a conducting two-dimensional electron gas (2DEG) with high mobility~\cite{a2,a3}, superconducting \cite{a4} and in some cases magnetic properties~\cite{a5,a6,a7}, and two-dimensional co-existence of both superconducting and magnetic properties~\cite{a8,a9}. In particular, the fundamental phenomenon that has generated a lot of interests is the LaAlO$_{3}$-thickness-dependent insulator-metal transition of LaAlO$_{3}$/SrTiO$_{3}$ interface. This phenomenon has several peculiar characteristics as shown by the various experimental observations summarized in Table I.

Recent high-energy reflectivity measurements have shown that in conducting LaAlO$_{3}$/SrTiO$_{3}$, a charge transfer of 0.5$e^-$ from LaAlO$_{3}$ into LaAlO$_{3}$/SrTiO$_{3}$ interface and oxygen vacancies have been observed \cite{a13}. While in insulating LaAlO$_{3}$/SrTiO$_{3}$, an intra-layer charge redistribution within the LaAlO$_{3}$ film has been found. However, a fundamental mechanism to explain properties in insulating, conducting, as well as an important \textit{step-function-shape} insulator-metal transition as a function of LaAlO$_{3}$ thickness in LaAlO$_{3}$/SrTiO$_{3}$  as one coherent picture is still lacking.

\begin{table}

  \setlength{\tabcolsep}{3.5pt}
  \renewcommand{\arraystretch}{1.3}
  \caption{A list of fundamental experimental observations from various experimental methods of transport, high-energy optical conductivity, and structural features.}
  \label{table:table1}
  \begin{center}
  \begin{tabular}{|m{0.75in}|m{2.45in}|}
  \hline
  \multicolumn{2}{|c|}{Observations for $<$ 4 unit cells (uc) of LaAlO$_{3}$}\\
  \hline
  Electrical transport \cite{a2,a3,a13} & 1. The samples were all insulating.\\
  \hline
  \multirow{4}{0.75in}{High-energy optical conductivity \cite{a13}} & 2. The samples were all insulating.\\
   & 3. A redistribution of 0.5$e^{-}$ from AlO$_{2}$ sub-layer to LaO sub-layer has occurred.\\
   & 4. There was no inter-layer charge transfer into the interface.\\
   & 5. No oxygen vacancies signature has been observed.\\
  \hline
  \multirow{3}{0.75in}{Structural features \cite{a15,a16}} & 6. An atomic buckling between cations and oxygen atoms was observed.\\
   & 7. These buckling decreased with increasing LaAlO$_{3}$ thickness.\\
   & 8. SrTiO$_{3}$ exhibited an opposite buckling at the near-interface region, which increased with increasing layer thickness.\\
  \hline
  \multicolumn{2}{|c|}{Observations for $\geq$ 4 unit cells (uc) of LaAlO$_{3}$}\\
  \hline
  \multirow{4}{0.75in}{Electrical transport \cite{a2,a3,a13}} & 9. The samples were all conducting.\\
   & 10. A sharp discontinuity of the electron gas density and insulator-metal transition at 4 uc LaAlO$_{3}$ has occurred.\\
   & 11. If LaAlO$_{3}$ thickness was further increased, the interface conductivity has remained relatively constant, independent of LaAlO$_{3}$ thickness.\\
   & 12. The surface of LaAlO$_{3}$ was always insulating.\\
  \hline
  \multirow{3}{0.75in}{High-energy optical conductivity \cite{a13}} & 13. A decrease of 0.5$e^{-}$ in LaAlO$_{3}$ film was observed.\\
   & 14. It was accompanied by an increase of 0.5$e^{-}$ at the interface, which resided in the SrTiO$_{3}$ side.\\
   & 15. Oxygen vacancies signature is observed in the LaAlO$_{3}$ film.\\
  \hline
  Structural features \cite{a15,a16} & 16. The buckling for the Al site for 5 uc LaAlO$_{3}$/SrTiO$_{3}$ film was collapsed.\\
  \hline
\end{tabular}
\end{center}
\end{table}

Previous theoretical studies \cite{a14,a36,a17,a18,a31,a22,a23,a24} discussed only some of experimental observations (in Table I). For instance, calculations based on perfect LaAlO$_{3}$/SrTiO$_{3}$ structures \cite{a17,a18} or buckling \cite{a14} did not describe at least two very important aspects observed in experiments. First, they did not explain the experimentally observed large discontinuity in the interface charge density as LaAlO$_{3}$ thickness increases above 4 unit cells (uc), but instead such calculations predicted this discontinuity only as a very gradual change. Second, the classical electronic reconstruction model and previous computational results would expect the surface of LaAlO$_{3}$/SrTiO$_{3}$ to also be metallic, which could strongly disrupt any conclusion drawn from transport and high-energy optical conductivity measurements regarding the interface charge density. Experimental observations instead show that the surface is insulating. Furthermore, recent first-principles calculations of a polarity-induced defect model \cite{a31} were not adequate to explain important observations such as internal charge redistribution \cite{a13} and lattice distortions \cite{a15,a16} of LaAlO$_{3}$ in insulating LaAlO$_{3}$/SrTiO$_{3}$.

Here, $via$ first-principles calculations, we propose a unified picture that is an interplay of electronic reconstructions (both between LaAlO$_{3}$ sub-layers, and between LaAlO$_{3}$ and SrTiO$_{3}$), lattice distortions, and surface oxygen vacancies compensates the polarization potential divergence introduced due to the polar nature of LaAlO$_{3}$, in different LaAlO$_{3}$/SrTiO$_{3}$ samples. Using this model, we are able to comprehensively explain the fundamental mechanism behind the observed physical properties (see Table I) as one coherent picture. Which of these mechanisms are dominant in a given thickness of LaAlO$_{3}$ are determined by the lowest total energy state in that particular set of conditions. We also show that our model can be applied to the LaAlO$_{3}$/SrTiO$_{3}$ prepared with different growth conditions, and other polar/nonpolar oxide interfaces.

\section {METHODOLOGY}
All calculations were performed by using density-functional theory based on Vienna ab initio Simulation Package (VASP) \cite{a34,a34b}  with the Perdew-Burke-Ernzerhof (PBE) approximation for the exchange-correlation functional and the frozen-core all-electron projector-augmented wave (PAW) method for the electron-ion interaction \cite{a34c}. The cutoff energy for the plane wave expansion was set to 400 eV. Gamma centered k-point grids for Brillouin zone sampling were set to 3$\times$3$\times$1 for ionic relaxations and 6$\times$6$\times$1 for static calculations, respectively. We used 2$\times$2$\times$1  LaAlO$_{3}$/SrTiO$_{3}$ supercells for both stoichiometric structures and the structures with oxygen vacancies, the latter of which were created by removing one oxygen atom at the surface AlO$_{2}$ sub-layer of the stoichiometric 2$\times$2$\times$1 LaAlO$_{3}$/SrTiO$_{3}$ supercells. The number of SrTiO$_{3}$ layers was fixed to 2 uc while the thickness of the LaAlO$_{3}$ film increased as 2, 3, 4, 5, and 6 uc. A much thicker (6uc) SrTiO$_{3}$ as substrate tested lead to smilar results reported here.  In order to minimize the interaction between neighbor surfaces, a 13 {\AA} vacuum was applied along (001) plane of all interface structures. The lateral lattice constant of these supercells were fixed to 7.896 {\AA}, twice as the equilibrium value of the bulk SrTiO$_{3}$, in good agreement with previous studies \cite{a22,a24}. All the atoms except the bottom SrO sub-layer were allowed to relax until the forces are smaller than 0.02 eV/{\AA}. Since SrTiO$_{3}$ is not a polar material, the dipole field, if any, must be quite small. To avoid the spurious electric field, dipole corrections were included. Calculations on the density of states of LaAlO$_{3}$/SrTiO$_{3}$ were performed, confirming that  there is no mid-gap state being introduced at the bottom of SrO sub-layer. The internal electric field in LaAlO$_{3}$ was calculated by the slope of macroscopically averaged electrostatic potentials \cite{a34d}.

The formation energy of the surface oxygen vacancies is defined by\cite{a25}:\[E_{f}=E_{total}-E_{host}+\mu(T,P), \] where $E_{total}$ is the calculated total energy of slabs with oxygen vacancies, $E_{host}$ the energy for the stoichiometric 2$\times$2$\times$1 supercells, and $\mu(T,P)$ is the chemical potential of oxygen, which is related with sample growth environment. Experimentally, the upper limit of oxygen partial pressure for layer-by-layer epitaxial growth of LaAlO$_{3}$ films on SrTiO$_{3}$ is 5$\times$10$^{-2}$ mbar, while the lower limit is 1$\times$10$^{-6}$ mbar at the growth temperature of 850 $^{\circ}$C \cite{a6}. With these data, we calculate the corresponding energy of the upper and lower oxygen chemical potential to be -6.77 eV and -7.18 eV, respectively.

\begin{figure}
\includegraphics [width=0.5\textwidth]{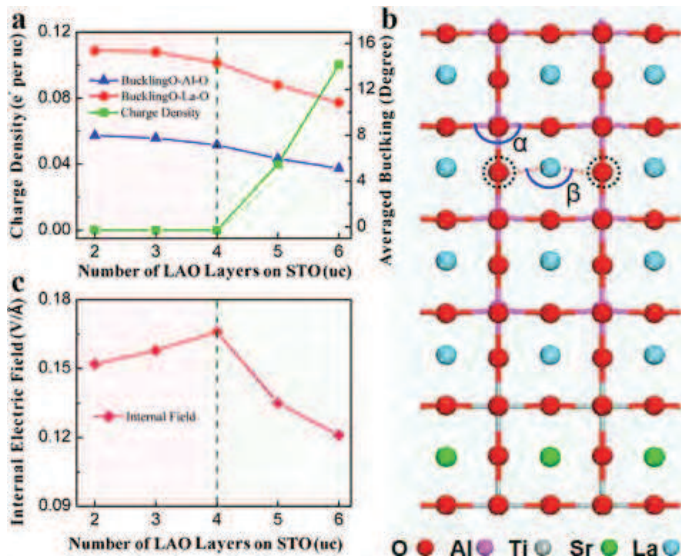}
\caption{Insulator-metal transition and lattice distortions in stoichiometric LaAlO$_{3}$/SrTiO$_{3}$. (a) LaAlO$_{3}$ thickness dependence of charge density at the interface, along with the average buckling of La-O-La and O-Al-O chains in LaAlO$_{3}$ films. To study the thickness-dependence behavior, we average the angles defined by 180$^{\circ}$$-$$\alpha$ and 180$^{\circ}$$-$$\beta$ for O-Al-O and O-La-O chains respectively, where $\alpha$ and $\beta$ are shown in Fig.1(b). (b) Structural guide for the buckling of O-Al-O and O-La-O chains. We should note that the circled oxygen atoms is not in the same plane with La atoms but we define their angle $\beta$ as the projected in-plane angle between their buckling. (c) The remnant electric field in LaAlO$_{3}$ films of LaAlO$_{3}$/SrTiO$_{3}$ at different LaAlO$_{3}$ thicknesses.}
\label{fig1}
\end{figure}

\section {RESULTS AND DISCUSSIONS}
To investigate the structural and electronic properties of stoichiometric 2, 3, 4, 5, and 6 uc LaAlO$_{3}$ layers on an SrTiO$_{3}$ substrate, we perform density functional theory (DFT) calculations on the thickness dependence of charge density at the interface, averaged lattice distortions of O-La-O and O-Al-O chains, and remnant internal electric field in LaAlO$_{3}$ layers. As shown in Fig.~\ref{fig1}a, there is an insulator-metal transition between 4 and 5 uc LaAlO$_{3}$/SrTiO$_{3}$. For the conducting cases, the charge density at the interfaces increases monotonically with {\textsl{d}}$_{LaAlO_{3}}$ . The lattice distortions between O and cationic atoms (as defined in Fig.~\ref{fig1}b) are found in LaAlO$_{3}$ film of all thicknesses. Overall, the lattice distortions of O-La-O chains are more pronounced than the lattice distortions of O-Al-O chains, consistent with experimental observations \cite{a15}. These cationic displacements introduce dipoles opposite to the polar electric field of LaAlO$_{3}$, and thus partially compensates it, leaving a remnant electric field in LaAlO$_{3}$.

Fig.~\ref{fig1}(c) shows the remnant electric field after partial compensation by the lattice distortions and electronic reconstructions. It can be seen that, below 5 uc of LaAlO$_{3}$, the lattice distortions cannot fully compensate the polar electric field, as the remnant electric field keeps increasing with the increase of {\textsl{d}}$_{LaAlO_{3}}$. When the {\textsl{d}}$_{LaAlO_{3}}$ $>$ 4 uc, a critical point is reached and the remnant electric field results in a potential difference that exceeds the LaAlO$_{3}$ band gap, which causes an internal Zener breakdown to occur. At this point, electronic reconstruction happens in LaAlO$_{3}$/SrTiO$_{3}$ via electrons transfer from surface AlO$_{2}$ sub-layer into interface TiO$_{2}$ sub-layers, leaving extra holes at surface AlO$_{2}$ and extra electrons at interface TiO$_{2}$ sub-layers, although the calculated charge density remains very small. We note that the phenomenon of electronic reconstruction was introduced in earlier study of polar surfaces in K$_3$C$_{60}$ \cite{a19}. This electronic reconstruction also compensates the polar electric field of LaAlO$_{3}$, leading to a sharp decrease of the remnant electric field after 4 uc of LaAlO$_{3}$ (see Fig.~\ref{fig1}c).

\begin{figure}
\includegraphics [width=0.45\textwidth]{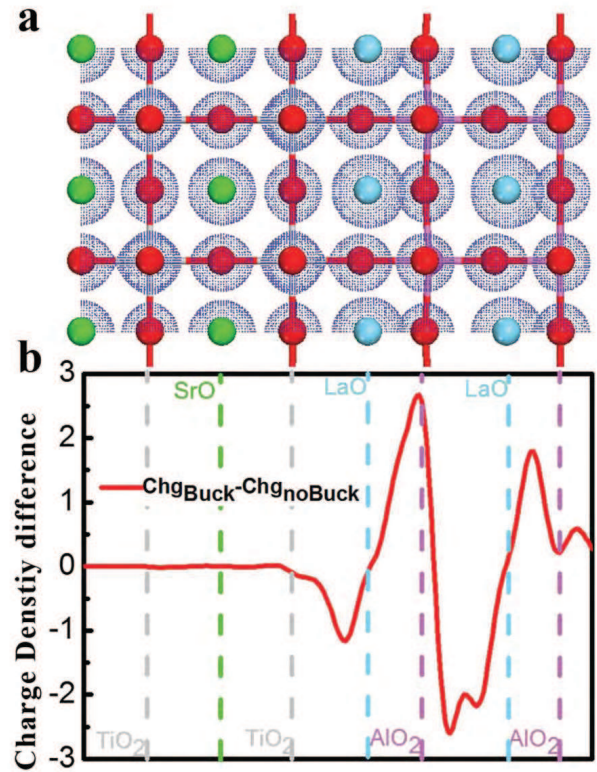}
\caption{In-plane averaged charge density of stoichiometric 2 uc LaAlO$_{3}$/SrTiO$_{3}$. (a) Structural configuration of stoichiometric 2 uc LaAlO$_{3}$/SrTiO$_{3}$ and the corresponding charge density distribution (isovalue = 0.027 e/{\AA}$^{3}$). (b) Difference of the xy-plane-averaged electron density between relaxed and unrelaxed cases.}
\label{fig2}
\end{figure}

Recent high-energy optical conductivity measurement of LaAlO$_{3}$/SrTiO$_{3}$ shows that there is charge redistribution between LaO and AlO$_2$ sub-layers in insulating LaAlO$_{3}$/SrTiO$_{3}$ \cite{a13}. In Fig.~\ref{fig2}, we compare the charge density distribution between relaxed and unrelaxed stoichiometric 2 uc LaAlO$_{3}$/SrTiO$_{3}$ structures and the internal charge redistribution between LaAlO$_{3}$ sub-layers is indeed found in our calculations. Because of the lattice distortions, the La atoms move upwards to the surface while the oxygen atoms in AlO$_2$ sub-layers move downwards to the interface (the La-O buckling dipole), making their orbitals more overlapped and thus more covalent than in the structures without lattice distortions. Although the overlap between Al atoms in AlO$_2$ sub-layers and the oxygen atoms in LaO sub-layers (the Al-O buckling dipole) counteracts the charge redistribution caused by the La-O buckling dipole, this effect is inferior due to the less pronounced lattice distortions of O-Al-O chains than the O-La-O chains. Since this internal LaAlO$_{3}$ charge redistribution has strong interplay with the lattice distortion effects, it also disappears once the atomic lattice distortions vanish. For this stoichiometric 2 uc LaAlO$_{3}$/SrTiO$_{3}$ structure, there is no charge redistribution between LaAlO$_{3}$ and SrTiO$_{3}$, consistent with experimental observations \cite{a13}.

\begin{figure}
\includegraphics [width=0.5\textwidth]{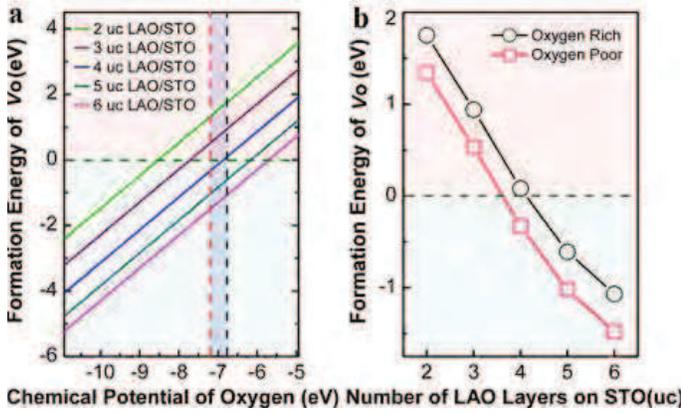}
\caption{Formation energy of surface oxygen vacancies in LaAlO$_{3}$/SrTiO$_{3}$ with different oxygen chemical potential and LaAlO$_{3}$ thicknesses. (a) Oxygen chemical potential dependence of formation energy of surface oxygen vacancies in LaAlO$_{3}$/SrTiO$_{3}$. The vertical black dash line indicates the oxygen rich condition, while the vertical red dash line indicates the oxygen poor condition. (b) The formation energy of surface oxygen vacancies for experimental oxygen-rich and oxygen-poor conditions as a function of LaAlO$_{3}$ thicknesses dependence in LaAlO$_{3}$/SrTiO$_{3}$.}
\label{fig3}
\end{figure}

Up to this point, we are able to explain to some extent the experimental observations of insulating LaAlO$_{3}$/SrTiO$_{3}$ (2 and 3 uc LaAlO$_{3}$/SrTiO$_{3}$), $i.e.$ the electronic redistribution inner LaAlO$_{3}$ sub-layers~\cite{a13}, the lattice distortions~\cite{a15,a16}, and the remnant electric field~\cite{a21}. However, this stoichiometric LaAlO$_{3}$/SrTiO$_{3}$ model is not able to fully explain the conducting LaAlO$_{3}$/SrTiO$_{3}$. First, the critical LaAlO$_{3}$ thickness happens at 5 uc instead of 4 uc as observed experimentally~\cite{a3}. Second, in conducting samples, the charge density increases with LaAlO$_{3}$ thickness, while experimentally the charge density of conducting LaAlO$_{3}$/SrTiO$_{3}$ is independent from LaAlO$_{3}$ thickness, as observed in the step-like insulator-metal transition~\cite{a3,a13}. Third, the charge density is much less than the 0.5e$^{-}$ observed experimentally~\cite{a13}. Fourth, the lattice distortions still remain after the interface becomes conducting, while experimentally~\cite{a15,a16} they collapse in conducting LaAlO$_{3}$/SrTiO$_{3}$.

Another important observation in high-energy optical conductivity of LaAlO$_{3}$/SrTiO$_{3}$ measurement was that the conducting LaAlO$_{3}$/SrTiO$_{3}$ exhibited \textsl{V}$_{O}$ signature in LaAlO$_{3}$.~\cite{a13} However, this signature was not observed in insulating LaAlO$_{3}$/SrTiO$_{3}$. This suggests that the formation conditions for \textsl{V}$_{O}$ in insulating and conducting LaAlO$_{3}$/SrTiO$_{3}$ are different. For this reason, we calculate formation energy (E$_{f}$) of \textsl{V}$_{O}$ in LaAlO$_{3}$ film of LaAlO$_{3}$/SrTiO$_{3}$ with different LaAlO$_{3}$ thicknesses. First of all, our calculations show that in LaAlO$_{3}$ film, the E$_{f}$ of \textsl{V}$_{O}$ is the lowest at the surface . Secondly, the E$_{f}$ of surface \textsl{V}$_{O}$ in LaAlO$_{3}$/SrTiO$_{3}$ also varies as a function of LaAlO$_{3}$ thickness and oxygen chemical potential, as shown in Fig.~\ref{fig3}. One clearly sees that E$_{f}$ of surface \textsl{V}$_{O}$ decreases as the thickness of LaAlO$_{3}$ increases, which means it is more energetically favorable to create surface \textsl{V}$_{O}$ in thicker LaAlO$_{3}$ film. Interestingly, when {\textsl{d}}$_{LaAlO_{3}}$ $\geq$ 4 uc, the formation energy of surface \textsl{V}$_{O}$ crosses through zero at most oxygen chemical potential window, indicating a high possibility to form the surface \textsl{V}$_{O}$.

\begin{figure}
\includegraphics [width=0.45\textwidth]{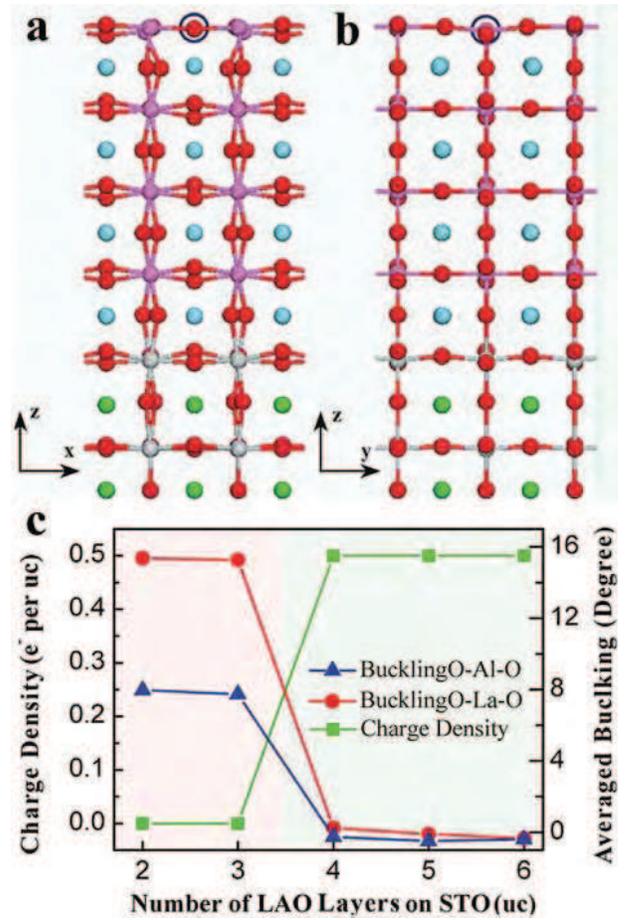}
\caption{Insulator-metal transition and lattice distortions in LaAlO$_{3}$/SrTiO$_{3}$, after the introduction of surface oxygen vacancies in conducting LaAlO$_{3}$/SrTiO$_{3}$. (a) Structural guide for the relaxed 4 uc LaAlO$_{3}$/SrTiO$_{3}$ along $<010>$ direction with surface oxygen vacancy. (b) Structural guide for the relaxed 4 uc LaAlO$_{3}$/SrTiO$_{3}$ along $<100>$ direction with surface oxygen vacancy. (c) LaAlO$_{3}$ thickness dependence of number of excess charge at the interface, and the average buckling of La-O-La and O-Al-O chains in LaAlO$_{3}$ thin films for different thicknesses of LaAlO$_{3}$ on SrTiO$_{3}$. The 2 and 3 uc LaAlO$_{3}$/SrTiO$_{3}$ are stoichiometric, while the 4, 5, and 6 uc LaAlO$_{3}$/SrTiO$_{3}$ are with surface oxygen vacancies.}
\label{fig4}
\end{figure}

Experimentally, the upper limit of oxygen partial pressure for layer-by-layer deposition of LaAlO$_{3}$/SrTiO$_{3}$ is~5$\times$10$^{-2}$~mbar with a typical deposition temperature of 850~$^{\circ}$C~\cite{a6}, implying this oxygen partial pressure is the oxygen-rich limit of LaAlO$_{3}$/SrTiO$_{3}$ deposition. The calculated formation energy of surface oxygen at this oxygen chemical potential energy is shown in Fig.~\ref{fig3}b. It can be seen that formation energy is equal to or below zero for {\textsl{d}}$_{LaAlO_{3}}$ $\geq$ 4 uc, even at this oxygen-rich limit, suggesting that the oxygen vacancies are energetically favorable to be formed in LaAlO$_{3}$ film with these thicknesses.

Based on this reason, we perform DFT calculations by including surface \textsl{V}$_{O}$ for 4, 5, and 6 uc LaAlO$_{3}$/SrTiO$_{3}$, which are experimentally found to be conducting~\cite{a3}. The relaxed LaAlO$_{3}$/SrTiO$_{3}$ structure after the inclusion of surface \textsl{V}$_{O}$ is shown in Fig.~\ref{fig4}a,b  along different directions. From here, the dipole-inducing lattice distortions as defined in Fig.~\ref{fig1}b vanish, consistent with experimental results~\cite{a15,a16}. The lattice distortions seen in Fig.~\ref{fig4}a and ~\ref{fig4}b are mainly caused by the absence of one surface oxygen atom in 2$\times$2$\times$1 supercell, and do not give rise to cationic dipoles.

\begin{figure}
\includegraphics [width=0.45\textwidth]{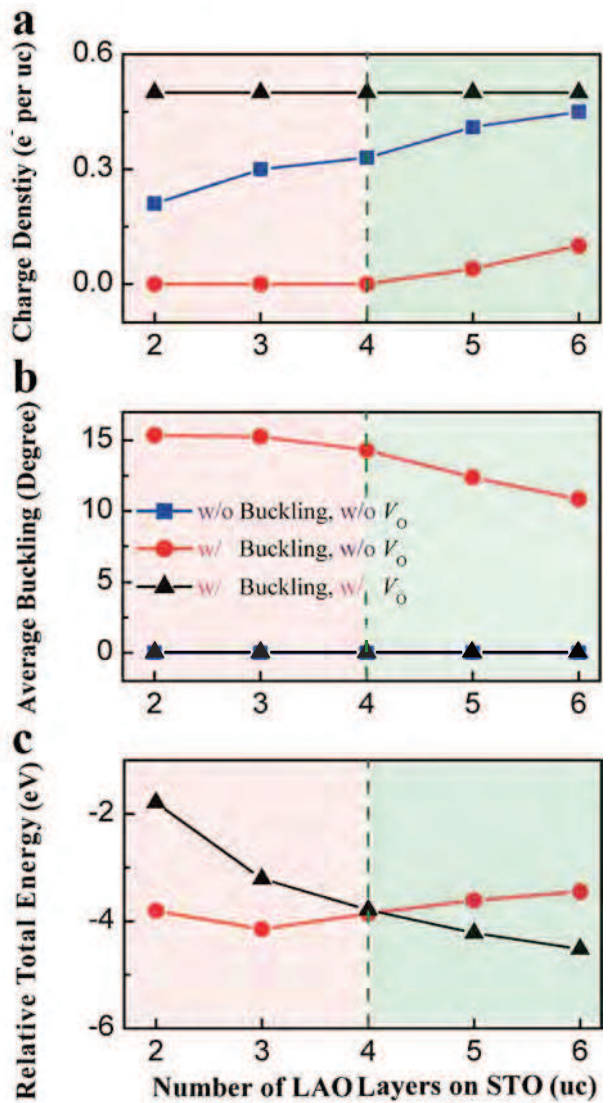}\\
\caption{A discontinuity phase-like transition due to interplay of electronic reconstructions, lattice distortions, and surface oxygen vacancies. Three cases of LaAlO$_{3}$/SrTiO$_{3}$ are considered: (1) LaAlO$_{3}$/SrTiO$_{3}$ without both lattice distortions and surface oxygen vacancies (blue squares), (2) LaAlO$_{3}$/SrTiO$_{3}$ with lattice distortion but without surface oxygen vacancies (red circles), and (3) LaAlO$_{3}$/SrTiO$_{3}$ with both lattice distortion and surface oxygen vacancies (black triangles). (a) Comparison of the interface charge density of each case. (b) Comparison of O-Al-O chains buckling in LaAlO$_{3}$ film of each case. (c) Comparison of relative total energy of the second and third cases, with the first case as the reference. Below 4 uc LaAlO$_{3}$ film (red-shaded region), the second case results in the lowest total energy and is consistent with experimental observations~\cite{a3,a13,a15}. On the other hand, above 4 uc LaAlO$_{3}$ film (green-shaded region) it is the third case that results in the lowest total energy and agreement with experimental observations~\cite{a3,a13,a15}.}
\label{fig5}
\end{figure}

The results of the calculations for the charge density are also shown in Fig.~\ref{fig4}c. (Note that the calculations for the charge density and buckling in 2 and 3 uc LaAlO$_{3}$/SrTiO$_{3}$ are performed without including surface \textsl{V}$_{O}$, the same as the results in Fig.~\ref{fig1}a.) It can be seen that the LaAlO$_{3}$ thickness dependence of the interface charge density now forms a step function, resulting in a discontinuity transition, consistent with experimental observations~\cite{a3,a13}. There are four main effects of surface \textsl{V}$_{O}$. First, the charge density is found to be 0.5e$^{-}$, which completely compensates the polar electric field, consistent with the experimental observation~\cite{a13}. Second, the amount of 0.5e$^{-}$ is consistent with the inter-layer charge transfer stipulated by the electronic reconstruction model. Third, the 0.5e$^{-}$ charge density is independent of LaAlO$_{3}$ thickness (and results in an insulating surface), also in good agreement with experimental results~\cite{a3,a13}. Fourth, we can reconcile the presence of surface \textsl{V}$_{O}$ with the reconstruction model for polar crystal terminations, as discussed below.

In the conventional electronic reconstruction model~\cite{a19}, without allowing for \textsl{V}$_{O}$ or the internal lattice distortions described above, the charge of 0.5e$^{-}$ at the interface is compensated by 0.5 holes in the mainly O-2$p$ band of the AlO$_{2}$-terminated surface of LaAlO$_{3}$. In this model not allowing for lattice distortions or other deviations the electronic reconstruction would occur immediately after even only one layer of LaAlO$_{3}$ was deposited (see Fig.~\ref{fig5}a), contrary with experimental observations~\cite{a3,a13}. Even more importantly, both the SrTiO$_{3}$ interface layer and the top AlO$_{2}$ surface layer would become metallic.
Allowing for the lattice distortions, as shown in Fig.~\ref{fig5}b, demonstrates that these internal distortions can compensate for the internal polarization potential divergence for {\textsl{d}}$_{LaAlO_{3}}$ $ <$ 4 uc, without electronic reconstruction into the interface. Meanwhile, for {\textsl{d}}$_{LaAlO_{3}}$ $\geq$ 4 uc the top of the valence band of LaAlO$_{3}$ and the bottom of the conduction band of SrTiO$_{3}$ starts to overlap. These result in both the interface and the surface becoming metallic with a very small but gradually increasing charge density of opposite sign as the LaAlO$_{3}$ thickness increases (see Fig.~\ref{fig1}a and ~\ref{fig5}a).

The energy to create \textsl{V}$_{O}$, however, approaches zero at 4 uc of LaAlO$_{3}$ (Fig.~\ref{fig3}) and becomes negative as LaAlO$_{3}$ thickness is further increased. In this case, the total energy of the system with 1/4 \textsl{V}$_{O}$ per AlO$_{2}$ sub-layer and reduced lattice distortions is lower than the total energy for the case without the surface \textsl{V}$_{O}$ and with the strong lattice distortions. Thus, this becomes the energetically favorable situation (see Fig.~\ref{fig5}c). This results in 0.5e$^{-}$ at the interface, while the top AlO$_{2}$ sub-layer remains insulating because the electrons left behind by the formation of \textsl{V}$_{O}$ are transferred to the interface. In this way the sample as a whole remains charge neutral as must be the case.

Based on these results, our model can naturally explain the LaAlO$_{3}$-thickness-dependent discontinuous transition at 4 uc LaAlO$_{3}$ of LaAlO$_{3}$/SrTiO$_{3}$. Below 4 uc of LaAlO$_{3}$, the internal polarization potential in LaAlO$_{3}$ is compensated by the lattice distortions of LaAlO$_{3}$ film without electronic reconstruction into the interface accompanied by the intra-layer charge redistribution within the LaAlO$_{3}$ film, resulting in insulating LaAlO$_{3}$/SrTiO$_{3}$. Above 4 uc of LaAlO$_{3}$, the internal polarization potential in LaAlO$_{3}$ is instead compensated by electronic reconstruction into the interface, stabilized by the energetically favorable spontaneous appearance of \textsl{V}$_{O}$ in the surface AlO$_{2}$ sub-layer of LaAlO$_{3}$, resulting in conducting LaAlO$_{3}$/SrTiO$_{3}$ interface with LaAlO$_{3}$-thickness-independent charge density.

Furthermore, we can classify three types of LaAlO$_{3}$/SrTiO$_{3}$ samples based on different experimental conditions \cite{a3,a33,a35,a32}. First, when the LaAlO$_{3}$ film is deposited on SrTiO$_{3}$ with moderate oxygen partial pressure, surface \textsl{V}$_{O}$ forms, which leads to the insulator to metal behavior of LaAlO$_{3}$/SrTiO$_{3}$ described in Fig.~\ref{fig4}c. Second, when this LaAlO$_{3}$/SrTiO$_{3}$ is post-annealed in high oxygen pressure, the experimentally-observed conductivity decreases, but is not completely gone \cite{a3,a35,a32}. Based on our model, this suggests that the oxygen annealing removes some of the surface \textsl{V}$_{O}$, leading to the insulator-metal transition behavior of stoichiometric LaAlO$_{3}$/SrTiO$_{3}$ described in Fig.~\ref{fig1}a with its much lower interface charge density. Third, when LaAlO$_{3}$ is deposited with very low oxygen pressure, it is experimentally observed that the conduction region of LaAlO$_{3}$/SrTiO$_{3}$ extends up to 500 $\mu$m \cite{a33}. This suggests that the low oxygen pressure condition generates a high number of \textsl{V}$_{O}$ in SrTiO$_{3}$ (instead of surface LaAlO$_{3}$), giving rise to three-dimensional electron gas at the interface \cite{a33}.

In conclusion, our result shows that interplay of electronic reconstructions, lattice distortions, and surface oxygen vacancies is responsible for the insulator-metal transition of LaAlO$_{3}$/SrTiO$_{3}$ with a step function at four LaAlO$_{3}$ unit cells and can be controlled by the sample growth conditions and environments.

\begin{acknowledgements}
We acknowledge Ilya Elfimov, Alex Zunger, and Matthew Sherburne for useful discussions and comments on teh manuscript. This work is supported by Singapore National Research Foundation under its Competitive Research Funding (NRF-CRP 8-2011-06 and NRF2008NRF-CRP002024), MOE-AcRF Tier-2 (MOE2010-T2-2-121), NUS-YIA, and FRC. We acknowledge the NUS Graphene Center and CSE-NUS computing centers for providing facilities for our numerical calculations. The work at UBC is supported by the Canadian funding agencies NSERC, and CRC as well as the University of British Columbia. 
\end{acknowledgements}

\end{document}